\documentclass[conference]{IEEEtran}
\IEEEoverridecommandlockouts
\usepackage{comment}
\usepackage{cite}
\usepackage{amsmath,amssymb,amsfonts}
\usepackage[colorlinks, citecolor=blue]{hyperref}
\usepackage{balance}
\usepackage{algorithm}
\usepackage{algorithmic}
\usepackage{graphicx}
\usepackage{textcomp}
\usepackage{xcolor}
\def\BibTeX{{\rm B\kern-.05em{\sc i\kern-.025em b}\kern-.08em
    T\kern-.1667em\lower.7ex\hbox{E}\kern-.125emX}}
\usepackage{balance}
    
\begin{document}

\title{HE2C: A Holistic Approach for Allocating Latency-Sensitive AI Tasks across Edge-Cloud\\

\thanks
}

\author{\IEEEauthorblockN{Minseo Kim, Mohsen Amini Salehi}
\IEEEauthorblockA{\textit{HPCC Lab, Computer Science and Engineering Department} \\
\textit{University of North Texas},
TX, United States \\
minseokim@my.unt.edu, mohsen.aminisalehi@unt.edu}
\and
\IEEEauthorblockN{Wei Shu}
\IEEEauthorblockA{\textit{Tenstorrent Inc}\\
 \textit{wshu2015@gmail.com}
}
}

\maketitle

\begin{abstract}
The high computational, memory, and energy demands of Deep Learning (DL) applications often exceed the capabilities of battery-powered edge devices, creating difficulties in meeting task deadlines and accuracy requirements.
Unlike previous solutions that optimize a single metric (e.g., accuracy or energy efficiency), HE2C framework is designed to holistically address the latency, memory, accuracy, throughput, and energy demands of DL applications across edge-cloud continuum, thereby, delivering a more comprehensive and effective user experience. HE2C comprises three key modules: (a) a ``feasibility-check module'' that evaluates the likelihood of meeting deadlines across both edge and cloud resources; (b) a ``resource allocation strategy'' that maximizes energy efficiency without sacrificing the inference accuracy; and (c) a ``rescue module'' that enhances throughput by leveraging approximate computing to trade accuracy for latency when necessary.
Our primary objective is to maximize system prolong battery lifespan, throughput, and accuracy while adhering to strict latency constraints.
Experimental evaluations in the context of wearable technologies for blind and visually impaired users demonstrate that HE2C significantly improves task throughput via completing a larger number of tasks within their specified deadlines, while preserving edge device battery and maintaining prediction accuracy with minimal latency impact. These results underscore HE2C’s potential as a robust solution for resource management in latency-sensitive, energy-constrained edge-to-cloud environments.
\end{abstract}


\section{Introduction}
The Internet of Things (IoT) can effectively support multiple deep learning (DL) applications simultaneously and continuously \cite{gubbi2013internet}.
To aid individuals with visual impairments in better understanding their surroundings, IoT-based systems utilize various DL applications, including neural networks for vision, sound, and text-based inference \cite{chang2020medglasses}.

IoT device tasks can be categorized as follows:
(a) Vision-based technologies, which include applications such as object detection and face recognition in images, content understanding in videos, and abnormal movement detection to mitigate potential hazards.
(b) Sound and text-based technologies, encompassing applications like ambient sound recognition, speech-to-speech translation, and text-to-speech conversion.
(c) Non-AI-based technologies, such as navigation systems, which operate independently of artificial intelligence.

In these systems, wearable devices capture sensor data, image or video frames, and audio streams.
The data is initially encoded and compressed before being transmitted to an edge-to-cloud (E2C) network.
Once received in the E2C continuum, the data streams are decoded and processed by cognitive modules using computationally intensive neural network models, such as those for object detection and text-to-speech translation.
The inputs are then organized into a set of DL tasks, each with a specified deadline.
The E2C continuum processes these tasks and provides feedback to the smart devices for guidance.

However, several challenges arise in implementing such systems practically.
One major issue is the battery life of edge devices.
The high computational and memory demands of DL applications can drain the batteries of these devices quickly.

Additionally, edge devices have limited computing and memory resources \cite{hong2019resource}.
Running multiple neural network (NN) tasks simultaneously and continuously can lead to resource exhaustion, limiting the throughput of tasks on edge devices.
These limitations can also lead to high task drop rates on edge-only systems, potentially impacting the user experience.

In this paper, we propose HE2C that consists of two main components:
(a) An ingress traffic analysis module that leverages both pre-analyzed statistics and real-time task parameters for each application.
(b) An E2C decision maker that intelligently distributes tasks between the edge and cloud servers, optimizing device energy usage and meeting task throughput and deadlines.
When edge resources are critically limited, the E2C decision maker activates a rescue module to ensure essential outputs for all tasks. The main contributions are as follows:

\begin{itemize}
    \item We present the development of a comprehensive edge-to-cloud resource allocation framework called HE2C, designed to account for the energy, memory, and processing limitations of edge devices, as well as the network latency associated with cloud access.
    \item We provide methods to maximize edge battery lifespan, throughput, and accuracy metrics while adhering to strict latency constraints.
    \item We analyze the experimental results and demonstrate that HE2C can significantly enhance the overall system throughput and extend the lifespan of edge energy while maintaining task latency constraints.
\end{itemize}

\section{Related Work}
Cloud-based systems face severe latency constraints as data volume and frequency increase, posing significant challenges to the performance and responsiveness of modern applications \cite{ali2020res}.
Consequently, edge computing has emerged as a complementary paradigm to cloud computing, mitigating these latency issues by bringing computation closer to the data source \cite{yousefpour2019all}.
This shift is particularly valuable for applications with stringent latency requirements, such as those in Internet of Things (IoT) ecosystems and Artificial Intelligence (AI) tasks.

The CODA framework \cite{zhao2020coda} effectively addresses resource contention during the training of multi-stage Deep Neural Networks (DNNs) by dynamically adjusting resource distribution based on current workload demands to enhance overall resource utilization.
However, CODA is less suited for latency-sensitive IoT applications, where tasks must be executed in real-time with minimal delays.
We propose a dynamic selection of the optimal execution environment either edge or cloud—based on comprehensive analytics of accuracy, energy consumption, inference time, throughput, durability, and sustainability.
This approach will avoid unacceptable delays, making it well-suited for latency-sensitive applications.

Mokhtari et al. proposed a framework called FELARE \cite{mokhtari2022felare}, which focuses on ensuring timely task completion while considering the constraints of limited energy resources.
While FELARE provides a solid foundation for energy-aware scheduling, it primarily emphasizes energy consumption and task completion rates.
In contrast, our work builds on this concept by incorporating a broader set of metrics into the decision-making process for task placement and execution. 

Zobaed et al. \cite{zobaed2022} proposed Edge-MultiAI, a framework for managing multi-tenancy of latency-sensitive deep learning applications on edge environments.
Their work focuses on resource allocation under multi-tenancy constraints, providing a strong foundation for addressing latency-sensitive workloads in heterogeneous systems.
HE2C builds on this foundation by incorporating additional considerations such as energy efficiency and task accuracy in edge-cloud environments.

Aslanpour et al. \cite{mohammad2024} introduced a performance-driven load balancing strategy tailored for heterogeneous serverless edge computing environments.
Their work emphasizes empirical weight tuning for optimizing various performance metrics, offering valuable insights into managing task allocation efficiently across diverse nodes.
In contrast, HE2C extends this approach by integrating a holistic view of latency, energy, and memory constraints for latency-sensitive AI tasks.

In our work, we build upon these foundations by further examining the trade-offs between energy consumption, task execution time, and overall system throughput, providing a more comprehensive approach to task allocation in edge-to-cloud environments.

\section{Proposed Solution}
\subsection{Architecture}

\begin{figure}[htbp]
    \centering
    \includegraphics[width=1\linewidth]{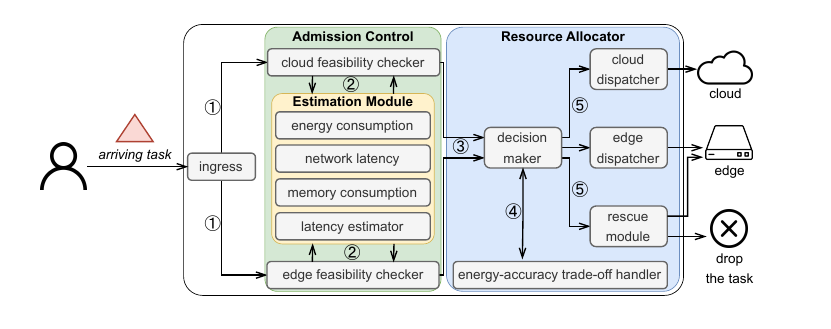}
    \vspace{-25pt}
    \caption{The architectural diagram illustrates the resource management system, incorporating feasibility checkers and a resource allocator.}
    \label{fig1}
\end{figure}

Fig.~\ref{fig1} illustrates a framework for efficient task management, integrating resources from edge devices to cloud infrastructure.
At its core, an admission control mechanism assesses task feasibility based on cloud and edge capacities, ensuring tasks are assigned to the appropriate resources.
Additionally, a real-time estimation module evaluates key performance metrics—energy use, latency, memory, and deadlines—continuously optimizing resource allocation to maximize performance and meet deadlines without sacrificing energy efficiency.

\vspace{-10pt}

\subsection{Admission Control}
The HE2C architecture features a robust admission control mechanism that performs feasibility checks for both cloud and edge servers, ensuring each task is assigned to the optimal environment.
This mechanism evaluates factors such as system capacity, energy consumption, latency, and task requirements to optimize performance and resource utilization.

HE2C follows a structured process flow for task allocation, starting with data profiling.
Metrics such as energy use, network latency, memory usage, and computational capacity are gathered from cloud and edge devices to assess the infrastructure’s current state and support workload generation.

The generated workload includes profiling data and task-specific parameters like deadlines, resource demands, and computational complexity.
This comprehensive view allows HE2C to evaluate the system’s capacity to handle new tasks without overloading any component.

This approach minimizes latency, reduces energy consumption, and maximizes resource utilization by selecting the environment best suited for each task based on factors like network congestion and real-time performance needs.

\subsubsection{Cloud Feasibility Checker}
The cloud feasibility check evaluates whether tasks can be executed in the cloud based on deadline constraints and energy availability.
The process begins by checking if the task’s deadline can be met, considering total cloud latency (data transmission, processing, and response time).
As shown in Alg.~\ref{alg1}, the algorithm checks if the latency exceeds the deadline, the task is deemed unfeasible for cloud execution \texttt{(Lines 6-7)}.
If latency is within limits, the algorithm then verifies energy availability, ensuring there is enough energy on the edge device for data transfer and cloud computation \texttt{(Lines 9-12)}.
If both conditions are satisfied, the task is approved for cloud execution.
If either latency or energy requirements are unmet, the task is rerouted to the edge device.
This approach ensures that only tasks meeting cloud resource limits are approved, optimizing system performance and energy efficiency.

\begin{algorithm}
\caption{Cloud feasibility checker}\label{alg1}
\begin{algorithmic}[1]
\STATE inputs: task $t_i$ with deadline $\delta_i$, edge energy E
\STATE $l_i \gets$ expected end-to-end latency of $t_i$ on Cloud
\STATE $\epsilon^u_i \gets$ estimate energy consumption to upload $t_i$ data to Cloud
\STATE $\epsilon^p_i \gets$ estimate energy consumption to get $t_i$ inference result from Cloud
\STATE $\epsilon^t_i \gets \epsilon^u_i + \epsilon^p_i$
\STATE {if}
    {$\delta_i < l_i$} then
    \STATE\quad {return infeasible on Cloud}
\STATE {else}
    \STATE\quad {if}
    {$E \geq \epsilon^t_i$} then
            \STATE\quad\quad\quad {return feasible on Cloud}
    \STATE\quad {else}
        \STATE\quad\quad {return infeasible on Cloud}
\end{algorithmic}
\end{algorithm}

\subsubsection{Edge Feasibility Checker}
The feasibility assessment for edge execution evaluates tasks based on deadline constraints, energy availability, and memory capacity.
As shown in Alg.~\ref{alg2}, the algorithm checks if the task’s deadline is shorter than the estimated edge execution time, accounting for any cold start delays \texttt{(Lines 5-6)}.
A cold start includes the time needed to load DL models or other resources.
If the execution time exceeds the deadline, the task is marked unfeasible for the edge.
If the deadline is met, the algorithm verifies energy availability on the edge device.
It checks if the battery power is sufficient for data upload and inference, neglecting the minimal energy needed for data upload but considering inference power \texttt{(Lines 8-9)}.
If energy is insufficient, the task is deemed unfeasible for edge execution.
Finally, if both deadline and energy criteria are satisfied, the algorithm checks if enough memory is available on the edge for DL applications, particularly under cold start conditions.
Insufficient memory would lead to performance issues or task failure, so tasks with high memory needs are redirected if necessary.
By ensuring tasks meet these criteria—deadline, energy, and memory—the algorithm optimizes edge resource usage and prevents overloads, balancing efficiency and resource management across the edge-cloud continuum.

\begin{algorithm}
\caption{Edge feasibility checker}\label{alg2}
\begin{algorithmic}[1]
\STATE inputs: task $t_i$ with deadline $\delta_i$,  edge energy $E$ and edge memory $M$
\STATE $c_i \gets$ expected completion time of $t_i$
\STATE $\epsilon^e_i \gets$ estimate energy consumption of $t_i$ inference on Edge
\STATE $\mu_i \gets$ maximum memory needed to run $t_i$ on Edge
\STATE {if}
    {$c_i \geq \delta_i$} then
    \STATE\quad {return infeasible on Edge}
\STATE {else}
    \STATE\quad {if}
    {$E > \epsilon^e_i$ AND $M > \mu_i$} then
                 \STATE\quad\quad\quad {return feasible on Edge}
        \STATE\quad\quad {else}
            \STATE\quad\quad\quad {return infeasible on Edge}
\end{algorithmic}
\end{algorithm}

\subsection{Resource Allocator}
The decision maker plays a key role in consolidating admission control results and selecting the optimal environment for task execution.
Based on feasibility checks from both the cloud and edge, it allocates tasks to the appropriate environment.

If either the cloud or edge feasibility check returns True, the decision maker instructs the relevant dispatcher to execute the task in that environment.
When both checks return True, it performs a further evaluation to determine the best environment, as detailed in Alg.~\ref{alg3}.
This process ensures that tasks are allocated based on energy consumption and accuracy requirements, enhancing system efficiency, performance, and energy utilization.

\noindent\textbf{Energy-Accuracy Trade-off Handler:} Decision maker employs a method, called energy-accuracy trade-off handler, (as shown in Figure~\ref{fig1}) to choose the execution environment such that both energy and accuracy metrics are fulfilled. For that purpose, we develop a method that operates based on the linear regression model \cite{montgomery2021}.
For an arriving task that is feasible on both Edge and Cloud, the trade-off handler considers the following metrics to decide whether energy or accuracy should be prioritized for that specific task.

\begin{itemize}
    \item task type ($t^t_i$): specifies the nature of the task, influencing the trade-off decision based on its sensitivity to accuracy or energy constraints.
    \item estimate energy consumption of $t_i$ on Cloud ($\epsilon^c_i$)
    \item estimate energy consumption of $t_i$ on Edge ($\epsilon^e_i$)
    \item accuracy of $t_i$ on Cloud ($\alpha^c_i$)
    \item accuracy of $t_i$ using the existing DL model on Edge memory ($\alpha^e_i$)
\end{itemize}

The handler uses linear regression to evaluate the trade-off between energy consumption and accuracy, determining whether to process each task on the cloud or edge.
This model-driven approach fine-tunes the balance between energy efficiency and accuracy, ideal for resource-constrained environments where priorities may shift.

The decision-making algorithm functions as follows:
It first checks if cloud execution uses less energy than edge execution \texttt{(Lines 6-7)}.
If so, the task is allocated to the cloud. If energy savings are not achieved in the cloud, the algorithm evaluates accuracy needs \texttt{(Lines 9-13)}.
When accuracy is a priority, the task is executed in the cloud to leverage its superior computational resources.

\begin{algorithm}
\caption{Decision making method for allocating a task across Edge-Cloud}\label{alg3}
\begin{algorithmic}[1]
\STATE inputs: task $t_i$, task type $t^t_i$, edge energy $E$
\STATE $\epsilon^c_i \gets$ estimate energy consumption of $t_i$ on Cloud
\STATE $\epsilon^e_i \gets$ estimate energy consumption of $t_i$ on Edge
\STATE $\alpha^c_i \gets$ accuracy of $t_i$ on Cloud
\STATE $\alpha^e_i \gets$ accuracy of $t_i$ using the existing DL model on Edge Memory
\STATE {if}
    {$\epsilon^c_i \leq \epsilon^e_i$} then
    \STATE\quad {dispatch $t_i$ to Cloud}
\STATE {else}
\STATE\quad decision $\gets$ call energy-accuracy trade-off handler ($t^t_i, \epsilon^e_i, \epsilon^c_i, \alpha^e_i, \alpha^c_i$)
    \STATE\quad {if the decision is Cloud then}
    \STATE\quad\quad {dispatch to Cloud}
        \STATE\quad {else}
    \STATE\quad\quad {dispatch to Edge}
\end{algorithmic}
\end{algorithm}

If neither energy consumption nor accuracy justifies cloud execution, the task is assigned to edge servers.
While edge servers may have slightly lower accuracy, they can meet requirements for tasks where energy consumption is less critical.

This multi-step approach ensures that each task is executed in the environment best suited for energy efficiency and accuracy, enhancing system performance and minimizing resource use.

\subsection{Rescue module}
In distributed computing, deciding whether a job can be executed on an edge server is essential, especially under resource constraints and strict deadlines. The decision to offload or drop a task is based on key criteria that evaluate both time and resource availability, as outlined in Alg.~\ref{alg4}.

This decision-making algorithm prioritizes cloud execution for tasks needing energy savings or high accuracy, while edge execution is preferred otherwise. This adaptive approach improves task management efficiency across distributed environments by dynamically responding to available resources and task requirements.

The rescue module applies two main checks to determine if a task should be processed on the edge or discarded. First, it checks if the task’s deadline is shorter than the estimated warm-start execution time on the edge (avoiding additional initialization). If the execution time exceeds the deadline, the task is marked as unfeasible for edge processing and dropped.
If the task meets the deadline, the second check evaluates the edge server’s remaining battery power, ensuring it can complete the task.
If energy is insufficient, the task is discarded to prevent overloading resource-constrained servers and conserve energy for higher-priority tasks.
By enforcing these deadline and energy checks, the algorithm ensures only feasible tasks are allocated to edge servers, thereby enhancing system stability and efficiency in managing distributed resources.

\begin{algorithm}
\caption{Rescue algorithm on Edge} \label{alg4}
\begin{algorithmic}[1]
\STATE inputs: task $t_i$ with deadline $\delta_i$, edge energy $E$ 
\STATE $c_i \gets$ expected completion time of $t_i$ on edge under warm start assumption
\STATE $\epsilon^e_i \gets$ estimate energy consumption for $t_i$ inference on Edge
\STATE if $\delta_i > c_i$ AND $\epsilon^e_i \leq E$ AND $t_i$ is in warm-start 
\STATE\quad dispatch $t_i$ to Edge
\STATE else
\STATE\quad drop $t_i$
\end{algorithmic}
\end{algorithm}

\section{Performance Evaluation}
\subsection{Experimental Setup and Evaluation Metrics}
To assess the effectiveness and performance of the HE2C system, we conducted a comprehensive set of benchmarks using four diverse DL applications: face recognition, text detection, text recognition, and image detection.
These applications were chosen for their varied computational demands and practical relevance, providing a robust foundation for evaluating system performance across different domains.

To make the evaluation as realistic as possible, we generated workload traces that reflect the true characteristics of these applications, including model size, inference accuracy, and computational complexity.
Using the E2C simulator~\cite{e2c2023}, we were able to recreate realistic edge-to-cloud execution scenarios, capturing the nuances of our experimental conditions.

HE2C is designed to manage resources efficiently within an edge-to-cloud environment, dynamically choosing the best execution environment for incoming tasks based on edge device conditions and task requirements.
For our assessment, we used a range of metrics to evaluate the system's impact on performance and QoS.

The experimental setup consisted of two main components:
(a) a Dell Inspiron, powered by an 11th Gen Intel(R) Core(TM) i5-1135G7 processor, which offers a balance of performance and energy efficiency, making it ideal for edge computing tasks, and
(b) the Chameleon Cloud platform, which served as the cloud computing resource, enabling us to simulate realistic edge-to-cloud task offloading and to assess the system's ability to manage distributed tasks.

This combination of a local edge device (Dell Inspiron) and a cloud platform (Chameleon Cloud) allowed for a thorough investigation of HE2C’s task allocation and resource management strategies across various workloads and conditions.

Overall, this experimental setup enabled us to rigorously test HE2C in real-world edge-cloud scenarios, offering valuable insights into its effectiveness in optimizing resource use, energy consumption, and QoS.

\subsection{Evaluating the impact of feasibility checker}
The objective of this experiment is to evaluate the effectiveness of the current feasibility checker in sustaining high task completion rates under diverse workload conditions.
By comparing system performance between the current feasibility checker and the single factor (latency) feasibility checker, this study aims to assess whether the multi-factor approach of the current checker effectively enhances system resilience, allowing a greater percentage of tasks to be completed on time as the workload increases.

\begin{figure}[htbp]
    \centering
    \includegraphics[width=0.7\linewidth]{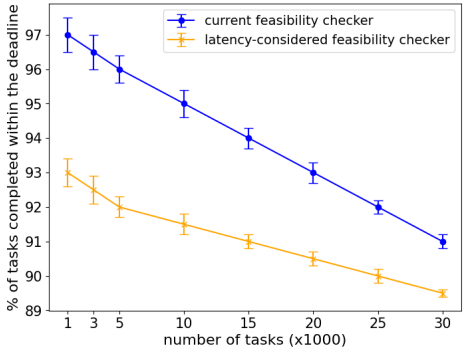}
    \vspace{-10pt}
    \caption{Impact of feasibility checkers on task completion rate across varying workloads}
    \label{fig2}
\end{figure}

As depicted in Fig.~\ref{fig2}, the current feasibility checker demonstrates a clear advantage, consistently achieving around 95\% task completion across all tested task volumes.
In contrast, the single-factor latency checker maintains a lower, fluctuating completion rate, averaging around 90--92\%.
These results underscore the current feasibility checker's essential role in bolstering system reliability and ensuring higher performance as task demand scales.

\vspace{-10pt}
\subsection{Evaluating the impact of trade-off handler}
The objective of this experiment is to evaluate the effectiveness of different trade-off handlers in balancing inference accuracy, energy consumption, and network latency for optimal system performance.
By analyzing the performance of these handlers across multiple metrics, this study aims to determine their ability to sustain high accuracy while minimizing energy and latency under varying conditions.

\begin{figure*}[htbp]
    \centering
    \includegraphics[width=0.9\linewidth]{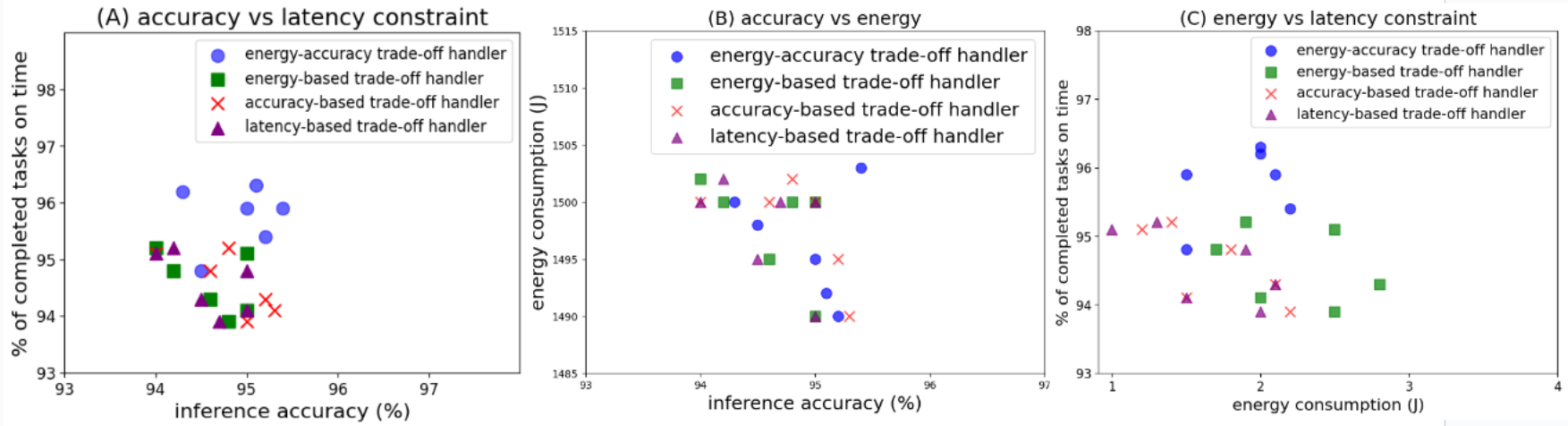}
    \vspace{-10pt}
    \caption{Comparative analysis of trade-off handlers across accuracy, energy, and latency metrics}
    \label{fig3}
\end{figure*}

As depicted in Fig.~\ref{fig3}, the energy-accuracy trade-off handler demonstrates a significant advantage, consistently maintaining inference accuracy around 94–97\% while achieving high task completion rates under latency constraints.
Similarly, the latency-based trade-off handler effectively prioritizes low latency while achieving comparable accuracy levels, albeit with slight variations in task completion rates.
In contrast, the energy-based trade-off handler shows higher latency variability but maintains accuracy within a similar range of 94–96\%.
Meanwhile, the accuracy-based trade-off handler prioritizes accuracy, with values approaching 97\%, but exhibits broader fluctuations in task completion under latency constraints.

As shown in Fig.~\ref{fig3}, the energy-accuracy trade-off handler achieves energy efficiency, with energy consumption values ranging between 1485–1510 J, while sustaining accuracy levels near 95\%.
The latency-based and accuracy-based trade-off handlers exhibit similar energy consumption ranges but show slight variations in accuracy and completion rates.
Conversely, the energy-based trade-off handler demonstrates higher energy consumption variability, suggesting its focus on energy efficiency comes at the cost of consistent performance.

Finally, Fig.~\ref{fig3} illustrates that the energy-accuracy trade-off handler excels in balancing energy consumption and latency.
It consistently achieves high task completion rates and low latency (approximately 1–2 J) across energy consumption values.
The latency-based trade-off handler performs similarly but with slightly higher variability in energy consumption.
The accuracy-based trade-off handler maintains consistent energy consumption but shows moderate latency (around 2–3 ms).
Meanwhile, the energy-based trade-off handler exhibits higher variability in latency, highlighting reduced optimization in this metric.

The energy-accuracy trade-off handler emerges as the most balanced solution, effectively optimizing accuracy, energy consumption, and latency.

\subsection{Evaluating the impact of rescue module}
The objective of this experiment is to assess the impact of the rescue module in sustaining high task completion rates under diverse workload conditions.
By contrasting system performance with and without the rescue module, this study aims to determine if the module effectively enhances system resilience, enabling a greater percentage of tasks to be completed on time as the workload increases. 

\begin{figure}[htbp]
    \centering
    \includegraphics[width=0.7\linewidth]{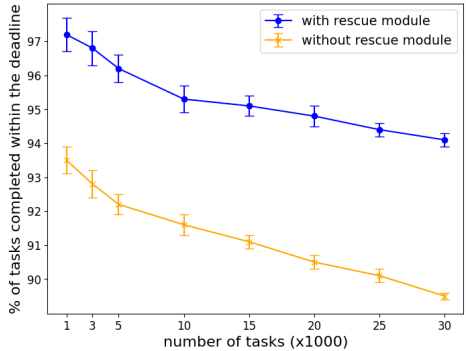}
    \vspace{-12pt}
    \caption{Impact of rescue module on improving on-time task completion rates across various number of tasks}
    \label{fig4}
\end{figure}

As depicted in the Fig.~\ref{fig4}, the rescue module exhibits a significant advantage, consistently achieving around 95\% task completion across all tested task volumes.
In contrast, systems operating without the module maintain a lower, fluctuating completion rate, averaging around 90--91\%.
These results underscore the rescue module’s essential role in bolstering system reliability and maintaining performance as demand scales.

\section{Conclusion}
In this research, we propose HE2C, a framework focused on optimizing resource allocation of concurrent latency-sensitive Deep Learning (DL) tasks across the edge-to-cloud computing continuum.
By considering longstanding challenges of energy efficiency, memory restrictions, and network latency, HE2C significantly enhances system performance and operational efficacy, particularly to enable ambient perception for the SmartSight project. 
HE2C integrates a comprehensive admission control mechanism with advanced decision-making algorithms to dynamically assess whether tasks should be executed on cloud servers or edge devices.
HE2C balances energy efficiency, memory utilization, and network latency while considering the task's latency constraints.
Experimental evaluations showed that HE2C significantly boosts system throughput, meets rigorous latency requirements, and maintains high inference accuracy, all of which are crucial for real-time AI applications.

There are several potential avenues for future work. One avenue is to consider the impact of latency-sensitive LLM-based applications. Another avenue is to consider urgent and non-urgent tasks in scenarios like SmartSight. For instance, for a blind and visually impaired user, obstacle detection is urgent and has to be executed within its time-limit, whereas, face detection and text detection are best-effort, i.e., there is no guarantee in their execution. A more interesting scenario is when the urgency of tasks are dynamic. For instance, in a building evacuation scenario, text detection can become urgent to assure helping user finding the ``exit'' signs. 

\section*{Acknowledgement}
We specially thank Ms. Bhanuja Ainary for her help and dedication in this work.
We would like to thank anonymous reviewers for their constructive feedback and Chameleon Cloud for providing us with resources for evaluation.
This project is supported by the National Science Foundation (NSF) through CNS Awards\# 2419588 and 2418188.

\end{document}